# The Status and Prospects of Phytoremediation of Heavy Metals

**Aniruddha Acharya[1*], Enrique Perez[2], Miller Maddox-Mandolini[1], Hania De La Fuente[2]**

[1]Division of Mathematics & Sciences, Delta State University, Cleveland, Mississippi, USA

[2]Department of Biology & Chemistry, Texas A&M International University, Laredo, TX, USA

**\* Correspondence:**
Aniruddha Acharya
aacharya@deltastate.edu & aniruddha1302@gmail.com
ORCiD - 0000-0002-6774-5934

**Abstract**

The release of heavy metals into the agricultural soil and waterbodies has been accelerated due to anthropogenic activities. They are not usually required for biological functions thus, their accumulation in biological system poses serious threat to health and environment globally. Phytoremediation offers a safe, inexpensive, and ecologically sustainable technique to clean habitats contaminated with heavy metals. Though several plants have been identified and used as a potential candidate for such phytoremediation, the technique is still at its formative stage and has been mostly confined to laboratory and greenhouses. However, recently several field studies have shown promising results that can propel large-scale implementation of this technology in industrial sites and urban agriculture. Realistically, the commercialization of this technique is possible if interdisciplinary approach is employed to increase its efficiency. This review presents a comprehensive narration of the status and future of the technique. It illustrates the concept of phytoremediation, the ecological and commercial benefits, and the types of phytoremediation. The candidate plants and factors that influences phytoremediation has been discussed. Finally, the physiological and molecular mechanism along with the future of the technique has been described.

## The problem of heavy metals

The term "heavy metal" has been poorly defined in numerous scientific literatures leading to confusion, however, the consensus is that they must have a density which is more than 5 g cm$^{-3}$ and an atomic number which is greater than 20. Though the metalloid arsenic and the non-metal selenium does not strictly adhere to the chemical definition of heavy metals but are widely placed in the category in literature related to environmental science and pollution due to their detrimental effects in living organisms (Ali & Khan 2018). In this review, we have clustered As and Se in the heavy metal category for ease of discussion and to imply their biological ramifications rather than chemical exclusivity. They have high atomic weight and most of them are toxic to biological system (Tchounwou et al. 2012; Fasani et al. 2018) though some of them are considered useful and necessary for completion of the life cycle of plants. Metals such as As, Ag, Cr, Cd, Pb, Hg and Se have no physiological functions in plants (Seth 2012) while others such as Fe, Cu, Zn and Mn are required in small amounts for optimal growth and development of plants (Cempel & Nikel 2006; Appenroth 2010; Asati et al. 2016). The accumulation of high amounts of these elements in agricultural soil, air, water, and human habitats is often referred as heavy metal pollution. Several natural and anthropogenic activities (Table 2) such as weathering of rocks, volcanic eruptions, mining, industrial and agricultural waste, fossil fuel use in household, transportation and industry, deposition of runoff water from agricultural fields to oceans, lakes, and other water bodies, oil spills in marine ecosystem and the military industry (Keshavarzi et al. 2015; Liu et al. 2018) are the source of such pollution. Several pesticides may contain compounds enriched in copper sulfate, lead arsenate besides the wide use of pesticides in agriculture contributes to heavy metal accumulation in arable lands (McLaughlin et al. 2000; Tchounwou et al. 2012; Missimer et al. 2018; Masindi & Muedi 2018).

Heavy metal pollution affects the health of many humans globally (Hembrom et al. 2020). Due to the fundamental chemical nature of elements, they cannot be further degraded by biological, chemical, or physical methods thus remains in the environment unaltered for long periods of time and their accumulation in air, water, and soil continues over a period due to industrialization, urbanization, anthropogenic activities, and disturbances in natural biogeochemical cycles thus posing environmental hazards and global concerns (Suman et al. 2018; Ashraf et al. 2019). They are taken up by plants and living organisms and thereafter travel to higher trophic levels through food chain due to consumption of contaminated water, air, and food. Higher plants can sequester them from soil or water into their tissues. Accumulation of these elements in the tissues of living organisms is called bioaccumulation. Their concentration can increase in tissues as they accumulate from lower to higher trophic levels in the ecosystem by a process called biomagnification (Ali & Khan 2019). One of the adverse effects of heavy metals in cell physiology is the production of free radicals or reactive oxygen species that can cause oxidative stress leading to damage of biomolecules, cell membrane, and cellular organelles (Ekmekci et al. 2008; Pandey & Madhuri 2014) leading to cell death (Popova et al. 2009; Xu et al. 2009). They adversely affect several biochemical, cellular, and physiological processes in plants that leads to decrease in relative water content in leaves and xylem vessels, reduced chlorophyll content, chloroplast damage, reduced photosynthetic and transpiration rate, decreased stomatal conductance, and decline in cell growth and enlargement (Ghori et al. 2019). The adverse effects of such pollutants on plant cell and physiology can lead to inhibition of plant growth (Popova et al. 2009; Xu et al. 2009). Thus, the accumulation of heavy metals in the environment such as agricultural soil and waterbodies is a serious health and environmental hazard. (Nguyen et al. 2019; Sandeep et al. 2019; Ali et al. 2021; Mitra et al. 2022).

**The concept and benefit of phytoremediation**

Plants can sequester ions elements from soil and store them in their cells at concentrations that are much more than compared to the concentration in their natural growing environment. Such accumulation of ions within cellular organelles and cytoplasm is an essential physiological process required to maintain metabolic activities and homeostasis because cells require several ions in certain minimal threshold concentrations to support optimal cellular metabolism (Rouached et al. 2010; Srivastava et al. 2020; Mulet t al. 2020; Yang et al. 2022). Such ions or elements are utilized in several metabolic pathways to maintain structural identity and functionality of cells (Dreyer 2021). Certain plants in nature have the capacity to accumulate and metabolize specific elements in their tissues in unusually high concentrations. Such high concentrations of the specific element concerned are not found in tissues of most naturally occurring plants in similar environment. These plants are called hyperaccumulators and certain hyperaccumulators can sequester specific elements. This relation between hyperaccumulators and the element it accumulates is highly specific both in terms of the plant species and the element it accumulates. (Van der Ent et al. 2013; Leitenmaier & Küpper 2013). There are several plants that has been identified in nature as hyperaccumulators and they can sequester metals from soil and water environments and translocate them to aboveground parts such as stems and leaves. The stem and leaf tissues of metal hyperaccumulators can accumulate metals at concentrations that are few magnitudes higher than the metal accumulation capacity of closely related species. Thus, such high concentrations are considered toxic or even lethal for most plants species but cannot cause any physiological damage to hyperaccumulating species (Table 1). Hyperaccumulators are often utilized to sequester specific elements with the target to clean the environment by a process called phytoremediation. The same property of hyperaccumulators is utilized to sequester commercially important elements such as gold, platinum and palladium by a process called phytomining. Thus, the process of phytomining and phytoremediation overlaps mechanistically but with different objectives. The use of plants in removal of hazardous compounds is referred as phytoremediation while the term bioremediation is used in a larger context where the use of any biological entity including plants for the removal of hazardous compounds from the environment is considered (Salt et al.1998; Pilon-Smits 2005; Ali et al. 2013; Wei et al. 2021; Shen et al. 2022 & Sharma et al. 2023). Thus, phytoaccumulation can be defined as a phytoremediation technique where harmful or commercially important metals can be sequestered by roots of hyperaccumulating plants and can be translocated to aboveground plant tissues where they accumulate over time.

On the contrary, phytostabilization is an emerging technology of phytoremediation which utilizes plants along with associated microbes in soil to prevent the spread and mitigate the harmful effects of pollutants in the environments by in situ treatments of contaminated soils, sediments, landfills, agricultural field, mines, military and industrial sites, household sewage, industrial and agricultural drainage water, groundwater, and air (Salt et al. 1995). Heavy metals from different wastes such as pesticides, herbicides, fertilizers, and effluents cannot be degraded, however, they can be immobilized, stabilized, and can be sequestered in plant tissues that can be further harvested (Dalvi & Bhalerao 2013). The primary goal of phytostabilization is to reduce the mobility of contaminants by sequestering them in roots or immobilizing them in rhizosphere through rhizospheric microbial processes. The process can be enhanced through addition of soil amendments and continuous soil monitoring (Bolan et al. 2011).

Besides phytostabilization of pollutants and their reduction from contaminated air, water, and soil ecosystems, the phytoremediation technology has several benefits, thus has emerged as a useful and popular technology over the past two decades. There are several direct and indirect benefits of this technology. It is a noninvasive alternative to complementary technology such as soil excavation, soil washing, and soil burning to decontaminate a polluted site. It is an economically feasible technology as is performed by autotrophs using the solar energy. It does not require the installation or maintenance of specialized instruments (Cameselle et al. 2019). The physical and chemical-based soil remediation methods, which mainly involve soil removal and burial at a prohibitive high price is not feasible on a long-term scale due to limited funds allocated by governments and several other agencies for environmental causes (Khalid et al. 2017; Liu et al. 2018). It is estimated that phytoremediation is less expensive than engineering-based remediation methods and is a financially manageable technique for long-term applicability (Wan et al. 2016). The process can be utilized to harvest metal rich plant tissues with high commercial values. The process of phytoremediation is efficient and versatile because several elements, inorganic and organic pollutants can be removed from solid, liquid, and gaseous substrates. The process can be applied for in situ recovery of vast areas of land, water, and air from pollutants and for a prolonged time. Despite large investments, the physical and chemical methods of remediation can induce secondary pollutant in the recovery site and causes significant changes in soil properties and soil microflora. On the other hand, the process of phytoremediation is ecofriendly and an efficient land management strategy as it reduces soil erosion, improves soil quality and organic content of soil (Wuana & Okieimen, 2011; Jacob et al., 2018). Finally, the process of phytoremediation can be of aesthetic value if ornamental plants or flowering plants are used in the process. The recent advances in molecular biology and gene editing along with the identification of several ion transporters can propel the advancement and commercialization of phytoremediation as a sustainable technology for heavy metal remediation (Kushwaha et al. 2015; Rai et al. 2020; Adeoye et al. 2022).

**Classification of phytoremediation**

Phytoremediation can be of several categories (Figure 1) based on the target material and their fate (Babu et al. 2021; Nedjimi 2021; Bhat et al. 2022). Many of these categories have significant overlap in their meaning and their mechanism of phytoremediation (Sytar et al. 2021). Here, we have categorized them mechanistically rather than being verbose.

Phytoaccumulation also referred as phytosequestration or phytoabsorption is a method of phytoremediation where toxic elements or compounds are sequestered from soil or water and translocated to the above-ground parts of plants (Ullah et al. 2011; Gul et al. 2022). Phytoextraction of phytominnning is the process of extracting metals of high commercial value from harvestable parts of hyperaccumulators. This is a commercially promising technology to mine high-value metals from polluted environments and the efficiency of the process depends on the chemical properties of soil, chemical state of the metal, and the species of plants used (Sinha et al. 2021).

Phytofiltration or rhizofiltration is the process of cleaning polluted water by plants or plant roots. Several elements, radioactive chemicals and other pollutants can be cleaned from waterbodies by this technique. Several aquatic plants or terrestrial plants with robust root system can be utilized for this process. However, high operational cost along with applicability of the process to only small volumes of water has been a major bottleneck for the large-scale use of this technique (Pang et al. 2023). Phytodesalination uses halophytic plants for cleaning of sites that are contaminated with metals and other pollutants. Halophytic plants are better adapted to saline environments as compared to their glycophytic counterparts and can be used to decontaminate saline soil and water environments (Li et al. 2019).

Phytostabilization or phytoimmobilization is the process of employing certain plants and their associated microbes to stabilize or reduce the bioavailability of pollutants in soil, however the process does not necessarily eliminate pollutants from soil. Certain elements can be made bioavailable by plant exudates and microbial activity that lowers soil pH in the rhizosphere. However, the leaching of such elements into ground water or their accumulation in food chain can be reduced through phytostabilization where the elements can be adsorbed, precipitated, complexed, and stabilized in the rhizosphere. Thus, transforming the toxic compounds to less toxic and preventing pollution by reducing their bioavailability (Rai et al. 2021).

Biotransformation is a process by which certain elements such as chromium, arsenic, and selenium are incorporated to organic compounds thus transforming them from their toxic chemical forms to less toxic organic compounds. Certain plants employ biotransformation either through organic conversion or through chemical reduction of these elements. An example of biotransformation is found in the species *Astragalus* where excess selenium is converted to nonprotein amino acids such as methylselenocysteine and selenocystathionine (Zhou et al. 2019; Wang et al. 2020). Phytodegradation is the enzymatic degradation of organic pollutants by plants and their associated microbes (Chlebek & Hupert-Kocurek 2019).

Phytovolatilization involves the use of plants to transform pollutants that are sequestered from soil into volatile forms and their subsequent transpiration to the atmosphere thus reducing their bioavailability. Organic and inorganic compounds can be converted into volatile forms and can diffuse through plant tissues before transpiring through leaves. The volatilization of selenium is thoroughly studied where selenium is converted to the amino acid selenocysteine or selenomethionine which can be methylated to a volatile compound called dimethylselenide and thus can transpire to the atmosphere (Pilon-Smits et al. 2014; Zayed et al. 2020). Phytovolatilization has been applied both naturally and through genetic engineering to remove mercury from soil and convert it into less toxic and volatile form. Genes such as merA and merB are widely studied for such processes, however, there are concerns raised about the safety of the technology (Leonard et al. 1998; Ruiz & Daniell 2009; Wang et al. 2012).

**Common candidates of phytoremediation**

Plants that can thrive on high metal concentration are called metallophytes and among them there are some plants that can not only survive but are capable of accumulating high concentration of metals. Such plants are classified as hyperaccumulators (Alford et al. 2010; Van der Ent et al. 2013). A large percentage of hyperaccumulators are represented by the Asteraceae, Brassicaceae, Cyperaceae, Euphorbiaceae, Fabaceae and Poaceae families of plants. There are nearly 500 species from 50 families that are categorized as hyperaccumulators. *Brassica juncea* and *Typha latifolia* L. has been utilized for phytoremediation of selenium while *Pteris vittate* is utilized to decontaminate Arsenic pollution. *Helianthus annuus* or sunflower is another good candidate for phytoremediation. Grasses such as *Trifolium alexandrinum* have been used for phytoremediation of heavy metals. Among aquatic plants, *Pistia stratiotes* or water lettuce has been used to decontaminate manganese pollution in waterbodies. Several aquatic plant species such as *Azolla*, water hyacinth, and duckweed are used for phytoremediation of waterbodies due their high growth rate, high biomass, adaptation capabilities, and can be easily harvested for target elements. Several plants (Table 1) including *Thlaspi*, *Brassica*, *Jatropha*, *Helianthus* and *Alyssum* have been investigated for their hyperaccumulation and phytoremediation properties thus are common candidates for phytoremediation. Genetically modified *Arabidopsis thaliana* and *Nicotiana tabacum* have been used for phytoremediation of mercury (Pollard et al. 2014; Baker et al. 2020; Reeves et al. 2021)

**Factors that influence phytoremediation**

An ideal candidate for phytoremediation must have several desirable characters, however, it is difficult to find a plant that has all those traits (Anton & Mathe-Gaspar 2005; Laghlimi et al. 2015; Shi et al. 2023).

Unless aquatic, the plants should have a deep and highly branched root system. Such a robust root system can interact with a large volume of soil and thus can sequester ions from them. The deeper root system will allow extraction of solutes from deeper regions of soil where the metal might have leached. However, if the candidate plant is aquatic the deeper and branched root system though important will not be as important a factor as in the case of terrestrial plants because waterbodies have more uniform gradient of ion concentration than the terrestrial environment. Thus, hyperaccumulating aquatic plants can sequester specific ions from contaminated water bodies even if they lack highly branched and deep root systems. The interaction of plant roots with the biotic and abiotic factors in the rhizosphere plays a major role in bioavailability of the elements. Thus, the nature of these interactions and soil chemistry are factors that needs to be considered before selecting a hyperaccumulator for in-situ phytoremediation (Chen et al. 2019; Oniosun et al. 2019).

Candidate plants for phytoremediation should have the capacity to tolerate high ionic concentrations in their growing environment. Toxic elements, pollutants and xenobiotic compounds can have detrimental effects on plant physiology even at small concentrations. Thus, an ideal candidate plant for phytoremediation must have the capacity to withstand high concentration of these elements in their habitat such as contaminated soil or water bodies (Pulford & Watson 2003). They must be able to sequester and compartmentalize the elements within their cell and maintain ionic homeostasis.

Besides the robust root system, another critical factor for an ideal candidate for phytoremediation will be the growth rate of the plant. A rapid growth rate is desirable so that the contaminated site can be cleaned in a relatively realistic time frame. Plants can take few months to years or decades to attain maturity and complete its life cycle. Some plants such as *Sequoiadendron* can grow for centuries. Phytoremediation directly depends on the physiology, growth, and development of the candidate plants. Biological processes such as growth and development are much slower than compared to chemical and mechanical processes. As phytoremediation is a biological process, so the decontamination of pollutants is much slower than different chemical and mechanical processes that can be used for the purpose. So, it is extremely important that the candidate plant selected for phytoremediation must have a high growth rate to achieve decontamination of polluted sites in a realistic timeframe. Biomass of the plant is another critical factor that needs to be considered when selecting candidates for phytoremediation. The amount of elements that can be extracted from contaminated sited by the process of phytoremediation can be calculated multiplying the concentration of the metal in harvestable parts of the hyperaccumulating plant to its dry mass. A plant having high biomass will thus be a more useful candidate than the one that has low biomass because the greater is the biomass, the higher will be the amount of the element that it can sequester from the contaminated site. Plant biomass is an abundant and renewable resource thus, a hyperaccumulating plant with high biomass can be repeatedly sowed and harvested to decontaminate land and waterbodies from pollutants and toxic metals (Rostami & Azhdarpoor 2019).

The ease of harvesting plant materials from a hyperaccumulating plant and the ease of cultivation and maintenance are few parameters that are important factors that needs to be considered while selecting plants as a phytoremediation candidate. If the plant parts accumulating pollutants are difficult to harvest then it raises the cost of phytoremediation, again if the pants are difficult to cultivate and needs high maintenance then the cost of the process also increases. A plant that is hardy, can grow in marginal soil and accumulates most of the sequestered pollutants in aerial organs such as stems, and leaves can serve as a good candidate plant for phytoremediation. However, if the plant is aquatic then harvesting the whole plant is relatively easier as compared to its terrestrial counterpart and in such cases the accumulation of pollutants in aerial parts is not a prerequisite because all parts of the plant could be harvested. The extraction of elements of high commercial value is one of the primary goals of phytoremediation. Thus, the process of phytoremediation will be economically competitive over physical and chemical methods of remediation if the extraction of elements from the harvested plant tissues is inexpensive and easily achievable.

Plants can get infected by a host of pathogens, thus the susceptibility of the candidate plant used for phytoremediation to different pathogens should be considered before cultivating it on a contaminated site. If the plant has a propensity for infection and cannot complete its lifecycle, then it will not be sustainable financially to employ it for phytoremediation. On an ecological perspective, caution should be employed such that the plants used for phytoremediation should not be a source of food for herbivores as it might lead to biomagnification of pollutants through food chain. This might lead to undesirable consequences of several diseases in human population and regional fauna. The genetic make-up and the physiological mechanism of the plant to efficiently sequester, translocate and accumulate toxic ions or detoxify pollutants is an important factor for phytoremediation. Unfortunately, most of the plants that can accumulate high levels of heavy metals have slow growth rate and low biomass. Plants that can grow on marginal soil and in limited water supply, are resistant to abiotic stress and can accumulate high amounts of toxic metals would be ideal for phytoremediation. Genetic engineering can be used to create transgenic lines with such properties. The use of hormones and chelating agents has been reported to augment phytoremediation; thus, a combination of physiological, chemical, and genetic techniques can be used to increase efficiency of the process (Razmi et al. 2021; Shen et al. 2022; Shi et al. 2023; Tan et al. 2023)

**Mechanisms involved**

Heavy metals are not natural to the ionomic composition of plants. They can cause severe disruption to the ionic homeostasis of the plant cell. Thus, plants generally employ strategies to avoid toxicity. Plants and microbes can interact in the rhizosphere to reduce the bioavailability of toxic ions or plants can compartmentalize them in organelles, cells and tissues that are not photosynthetic in nature. On the contrary, hyperaccumulating plants have evolved to withstand high concentration of metals in their tissue and they employ several strategies that includes, chemical, biochemical, microbial, cellular, and molecular mechanisms to manifest such objectives (Memon & Schröder 2009; Zhang et al. 2018; Kushwaha et al. 2018; Awa & Hadibarata 2020).

Chemical chelating agents such as EDTA when applied to soil enhances the phytoremediation of heavy metals. The chelating agents form metal-chelate complex that is transported to different tissues in plant where they get accumulated. Besides the chemical chelating agents that can be applied externally to the soil, plants can produce endogenous compounds that can chelate metals (Dipu et al. 2021; You et al. 2022). Phytosiderophores and peptides such as phytochelatins and metallothioneins are biochemical chelating agents that can chelate ions and can produced in response to metal deficiency. The relation between metal chelating peptides and metals are highly specific and such interactions helps to reduce phytotoxicity and maintain cellular homeostasis (Raskin et al. 1997; Pilon-Smits & Pilon 2002; Ruiz et al. 2011; Török et al. 2015, Sharma et al.2023). Several metabolites, metal-proteins, chelating compounds, and enzymes such as metallothioneins, phytochelatins, glutathione-S-transferases and phytosiderophores have been utilized to enhance the process of phytoremediation (Pilon-Smits & Pilon 2002; Memon & Schröder 2009; Sharma et al. 2023; Narayanan & Ma 2023; Thakuria et al. 2023). Metallothioneins are low molecular weight, cysteine-rich, cytosolic proteins that can bind to several metals such as cadmium, copper, zinc, and arsenic, thus can reduce their harmful effects in plant and animal cells through chelation, sequestration, detoxification, and metal homeostasis (Goldsbrough, 2000; Hall 2002; Guo et al. 2008). Genes encoding metallothioneins have been identified in several plants including *Arabidopsis*, sugarcane, and rice, while metallothionein genes has been overexpressed in plants such as tobacco and *Arabidopsis* resulting in heavy metal tolerance (Misra & Gedamu, 1989; Evans et al., 1992; Pan et al., 1994; Joshi et al., 2016). Metallothioneins are divided into several sub-groups based on the position of cysteine residue in their polypeptide and are increasingly becoming and important area of research in the context of phytoremediation (Zou et al., 2022). Phytochelatins are enzymatically synthesized low molecular weight polypeptides that are involved mostly in metal detoxification, metal homeostasis and abiotic stress. Phytochelatins are found in plants, animals, and fungi (Guo et al. 2008; Chatterjee et al. 2020; Chia 2021). They are rich in cysteine and are structurally related to glutathione synthetase. There are several structural variations of phytochelatins, however, they chelate metals through their thiol groups and are synthesized by the enzyme phytochelatin synthase (Cobbett 2000; Ovečka and Takáč, 2014). The overexpression of enzymes related to phytochelatin biosynthesis in *Brassica* was reported to express higher levels of phytochelatins in the transgenic plant tissues resulting in higher cadmium tolerance of the plant (Zhu et al. 1999). Glutathione-S-transferases are involved in phytochelatin biosynthesis and have been linked with copper, aluminum, and arsenic tolerance in plants (Ezaki et al. 2000; Maughan and Foyer 2006; Tiwari et al. 2022). Besides, heavy metal tolerance and detoxification, they have been reported to confer defense against oxidative stress in plants and yeast due to their thiol group (Grant et al. 1996; Vestergaard et al. 2008). Glutathione-S-transferases has been reported to have role in modulating gene expression (Foyer et al. 2001). Phytosiderophores are low molecular weight metabolites released by roots in response to iron deficiency and supports plant growth in iron-deficit soils. Phytosiderophoes can chelate iron molecules, however they have also been reported to chelate toxic elements and thus can be an ideal candidate to augment phytoremediation (Takahashi et al., 2001; Puschenreiter et al. 2017).

Ions in soil solution enters the plant root cells through membrane-bound transporters. The soil solution encounters barriers such as exodermis and endodermis before reaching vascular tissues. The metal can be stored in roots or can be transported to above-ground parts of plant through vascular tissues. To maintain homeostasis most heavy metals are transported to vacuoles and thus are compartmentalized within a cell. The soil surrounding roots are sites of increased microbial activity. Such activity results in enzymatic reactions that can detoxify pollutants and reduce the bioavailability of toxic ions. The plant growth promoting bacteria or PGPR can also influence the reduction of heavy metal stress in plants by enhancing nitrogen fixation and phosphorous absorption by roots (Gulzar & Mazumder 2022).

Identification of the molecular machinery involved in acquisition, transport, storage, and metabolism of heavy metals might lead to the understanding of the detailed mechanism of phytoremediation. Such, investigations might accelerate the development of transgenic plant lines that might be more efficient in phytoremediation. Several proteins such as ZIP (ZRT-IRT-like proteins), NRAMP (naturally resistant associated macrophage protein), HMA (heavy metals P1B-type ATPases) and MTP (metal tolerance protein) that belongs to the category of CDF protein (cation diffusion facilitator) have been associated with heavy metal tolerance of plants. Increased expression of ZIP proteins was associated with higher acquisition, transport, and accumulation of Zn in plants while the NRAMP proteins were localized in plasma membrane and tonoplast and were associated with Fe transport in plants. The HMA proteins were localized in plasma membrane, vacuole and chloroplast and were associated with heavy metal homeostasis (Yuan et al. 2012; Chaudhary et al. 2016; Zhang et al. 2017; Tian et al. 2021)

Members of the ZIP transporter family such as ZRT-IRT like protein are involved in Zn and Cd transport. Gene expression studies involving several genes of the ZIP and IRT like proteins confirmed their role in Zn and Cd transport in plants (Yang et al. 2009; Zhang et al. 2013). In rice, the tissue-specific localization of ZIP transporters such as *OsZIP5* and *OsZIP9* and their role in Zn and Cd uptake has been confirmed though mutation experiments (Lee et al.

2010; Yang et al. 2020). Genome-wide identification and transcriptome analysis of genes of HMA family indicated their role in Cd stress in plants (Ye et al. 2022). Cadmium resistance in plants has been attributed to the high transcription levels of genes of the UPS pathway (ubiquitin proteosome system) that might assist in degradation of misfolded protein (Shu et al. 2019). Arsenic resistance in plants has been related to transmembrane transporters such as ACR3 (arsenical compound resistance 3) while arsenic transport is related to arsenite effluxer such as PvAsE1 (Ali et al. 2012; Yan et al. 2022). The role of several high-affinity Pi transporters in As transport in plants has been confirmed (Meharg & Macnair 1992; Sun et al. 2019). Nickel hyperaccumulation and their sequestration in vacuoles of plant cells is attributed to IREG/Ferroportin transporters (García et al. 2021) and NcIREG2 transporter (Nishida et l. 2020) while transporters such as OsNramp5 and OsMTP9 are related to Mn transport in plants, other transporters and proteins involved in Mn transport are Nramp5 and MTP8.1 (Ishimaru et al. 2012; Chen et al. 2013; Ueno et al. 2015). The role of the gene *OsMYB45* has been identified and it confers cadmium resistance in rice (Hu et al. 2017).

**Recent field-studies on heavy metal phytoremediation.**

In the past few years, several field experiments have yielded promising results on heavy metal recovery from contaminated sites through the application of phytoremediation technology. In southern China the phytoremediation potential of chicory plant was investigated for cadmium recovery from contaminated fields. The effect of crop rotation involving rice and chicory plants were also estimated over a period of seven seasons resulting in a comprehensive study over a period of four years. The exhaustive dataset resulting from such studies indicated that an amount of 407 gram of cadmium per hectare were extracted from contaminated paddy fields after a period of four years (Deng et al. 2023). In another field study, chicory plants were grown in contaminated agricultural soil to estimate their efficacy for cadmium phytoremediation. High amount of cadmium was accumulated in chicory leaves and an average of 320 gram of cadmium per hectare were extracted from the contaminated site, thus indicating the phytoremediation potential of chicory plants for cadmium recovery from contaminated agricultural fields (Wu et al. 2023). Field experiments involving estimation of cadmium recovery by *Sesbania cannabina* when intercropped with rice indicated that the plant can be used to reduce cadmium levels in contaminated paddy field which in turn reduces cadmium levels in rice grains to 0.18 mg $Kg^{-1}$, an amount that is considered safe for consumption by food safety standards in China (Kang et al. 2021). Water-soluble chitosan enhanced phytoextraction of cadmium by *Hylotelephium spectabile* from contaminated soil and was concluded to be an effective soil amendment for cadmium extraction in fields (Guo et al.

2020). Cadmium and zinc polluted farmlands were utilized as experimentation site to investigate *Chrysanthemum indicum* L. as a candidate plant for phytoremediation. The three-year study concluded that the plants were able to reduce 78.1% of cadmium and 28.4% of zinc from the contaminated sites (Luo et al. 2020). In another field study, several cultivars of *Brassica juncea* L. were identified as good candidate for cadmium and lead removal from low o moderately contaminated soils (Gurajala et al. 2019). Field studies involving *Pennisetum* species identified it as an ideal candidate for cadmium and copper removal from contaminated fields (Xu et al. 2019). Plants such as sunflower and sesame are known to have high tolerance for metals thus, the synergistic effect of crop rotation involving sunflower and sesame was performed in field conditions to estimate their cumulative effect of heavy metal phytoremediation from soil. Several elements such as cadmium, copper, lead, and zinc were successfully recovered from soil by the phytoremediation process involving sunflower and sesame (Zhou et al. 2020). Phytoremediation experiments involving field studies indicated that sorghum was an ideal candidate as compared to maize and *Atriplex* for phytoremediation of cadmium, copper lead, and zinc from contaminated soil (Eissa & Almaroai 2019). Giant miscanthus were utilized for phytoremediation of copper, lead and zinc from abandoned flotation tailings near Serbia, results indicated that the miscanthus variety was capable of accumulating high amount of copper, lead and zinc in their roots (Andrejic et al. 2019). In another multi-year field-study in Croatia involving giant miscanthus, it was found that the plant can effectively accumulate copper, zinc, arsenic, strontium, manganese, titanium, iron, and molybdenum, thus can be used for phytoremediation from contaminated sites (Pidlisnyuk et al. 2020). A two-year controlled study followed by a three-year in situ experiment indicated that ferns such as *Pteris vittata* L. can be an ideal candidate for phytoremediation of arsenic from contaminated industrial sites. The aerial parts of the plants recorded a high amount such as 750 mg Kg$^{-1}$ of arsenic accumulation (Cantamessa et al. 2020). Industrial landfills were utilized as experimental sites to estimate the phytoremediation capacity of *Salix* species in combination with arbuscular mycorrhizal fungi. Several elements such as cadmium, zinc, copper, lead, tin and barium were recovered, or their concentrations were found to be reduced from contaminated soil indicating *Salix* to be an ideal candidate for heavy metal phytoremediation (Dagher et al. 2020). Finally, a recent study investigated the potential of grasses such as *Vetiveria zizanioides* in the removal of toxic metals from industrial wastewater fed soils. Results indicated that several metals such as copper, lead, iron, cadmium, and zinc could be successfully removed from contaminated soil thus making it an ideal candidate for development of green space in urban and industrial area (Kafil et al, 2019). Besides the terrestrial field studies, constructed wetlands offers a nature-based solution for phytoremediation of heavy metals and is increasingly getting popular due to encouraging results. Compared to terrestrial environment, the aqueous environment allows phytoremediation of a greater volume of pollutants thus being more cost-effective. The diverse population of macrophytes and microbial consortia can improve efficiency in treating wastewater and pollutants from different sources such as hospitals, domestic waste and industrial effluents. Several studies indicate the potential of constructed wetlands for large-scale phytoremediation of pharmaceuticals, organic volatiles, toxins, pollutants including heavy metals such as chromium and mercury (Sukumaran 2013; Gomes et al. 2014; Herath & Vithanage 2015; Men & Ghazi 2018; Sharma et al. 2022).

**Use of genetic engineering in phytoremediation**

The CRISPR technology for genome editing is one of the latest advancements in recombinant DNA technology that supersedes the power and precision of previously known genetic engineering technologies while reducing the time and cost for generation of mutants and transgenic lines. The significant reduction in DNA sequencing costs and the availability of genome sequence for several plant species has accelerated the use of CRISPR in the modification of plant genome (Figure 2) using genes from distant species or modulating the expression of endogenous genes (Chen et al. 2019). Several mechanisms involved in heavy metal phytoremediation such as stabilization, sequestration, translocation, accumulation, and transformation can be enhanced through this technology (Venegas-Rioseco et al. 2021). Genes encoding metal binding proteins such as metallothioneins; metal transporters such as ZIP, MATE, HMA and metal chelators such as phytochelatins has been the primary targets to enhance phytoremediation through CRISPR technology. However, manipulation of hormones, root exudates and pathways involved in oxidative stress has been targeted for detoxification, stress regulation and phytoremediation (Koźmińska et al. 2018; Talukder et al. 2023). Activation, repression, and overexpression of genes are the common mechanisms employed for controlling gene expression (Wu et al. 2010; YAN et al. 2023). However, it is essential to manipulate several genes of a pathway to obtain desired results in phytoremediation because most of the phytoremediation mechanisms involves a cassette of genes that works in tandem. Thus, the focus should be on manipulating key genes in a pathway rather than focusing on a single gene and CRISPR allows such manipulation thus is increasingly becoming the primary technology for genetic engineering in plants. CRISPR has been utilized towards the direction of enhancing phytoremediation in rice, corn, and poplar with promising results (Tang et al. 2017; Agarwal & Rani 2022). The technology can be used to create transgenic lines with higher tolerance for toxic ions, hyperaccumulation and phytoremediation capacity while gene pyramiding can help in increasing biomass and accelerating growth of such hyperaccumulators (Bhattacharyya et al. 2023). Finally, care must be taken to eliminate ecological risks of gene transfer to native flora and biomagnification of heavy metals in native fauna. Field trials of CRISPR mediated transgenic lines are necessary to gather conclusive data on the safety and wide-scale applicability of this technology.

**Challenges and future perspective**

The fundamental understanding of the biological process of phytoremediation at a biochemical and molecular level is essential for the development of genetically engineered hyperaccumulators that can be utilized for phytoremediation. Currently, the technology is at its infancy and though many plants have been identified as hyperaccumulators, most of the studies are limited to superficial analysis of the uptake of heavy metals by a plant species (Nouri et al. 2009; Hauptvogl et al. 2019; Antoniadis et al. 2021). Quantitative field study to estimate the commercial capacity of the technique and studies involving genetic investigation of the metabolic pathways that allows hyperaccumulation are essential for scale-up of this technology to make it a realistic tool for environmental sustainability and economic benefits. The uptake, translocation, compartmentation, and conversion of metal pollutants to less harmful compounds and their fate has been described to a considerable extent, however, the mechanisms involved have not been dissected. The role of the microbes dwelling in the rhizosphere and assisting the hyperaccumulators in heavy metal uptake is reported but the significant and complex interactions that are involved in the process needs further investigations (Sharma et al. 2021). The mechanisms that allow the biological system of hyperaccumulators to tolerate such elevated levels of heavy metals might have evolutionary origins by which they were able to protect themselves from pathogens and herbivores (Ernst et al. 1990). The ecological implications of growing hyperaccumulators in large scale is also important to understand because if the plant tissues enter food chain through consumptions of plant parts by herbivores, then it might become a health hazard in the long term (Ali & Khan 2019).

Future development of the process of phytoremediation is dependent on the detailed understanding of the role of microbes, membrane transporters, genetic and metabolic networks besides estimation of their efficacy in field and investigating the long-term ecological implications. The rhizosphere is a site of interactions between plant roots and microbes and a significant relationship exists as a large amount of photosynthates are exudated from the roots to the region that is utilized by the dwelling microbes. Thus, it is extremely important to understand the role of microbes in the phytoremediation process (Agarwal et al. 2020). The role of membrane bound transporters in phytoremediation is one of the aspects that needs rigorous studies. Though several ion transporters have been identified and their role has been established, the transporters specifically involved in phytoremediation is not well studied. Since these metals are not required by the plant to complete their life cycle and in some cases are required in a much smaller amount. Thus, their accumulation might be a byproduct of sequestration of a similar ion that matches with their chemical character. Due to chemical similarity, ions can be transported by the same membrane-bound transporters such as calcium/strontium (Jovanović et al. 2021) and potassium/rubidium (D'Hertefeldt & Falkengren-Grerup 2002). Thus, it is important to understand the nature and route of transport. The tissue or cell specific distribution and activity of these transporters can illustrate the pattern of ion accumulation. Recent studies indicate the distribution of ions in seedling roots are tissue specific (Pesacreta & Hasenstein 2018; Pesacreta et al. 2021; Acharya & Pesacreta 2022; Acharya 2022; Acharya & Pesacreta 2023). Anatomical structures such as Casparian strip and chemical complexes such as lignin and suberin plays an important role in ion transport regulation, thus, their role in the perspective of phytoremediation and ion transport might be a pertinent topic of investigation. It is important to understand the role of different ion-transport boundaries in apoplastic and symplastic route of heavy metal translocation.

The development of genetic tools, gene editing technology and sequencing of nucleic acids have allowed to understand the metabolic and regulatory network of several biological processes in plants, animals, and bacteria. Identification of novel genes and enzymatic pathways that confers heavy metal tolerance to hyperaccumulators is important to create transgenic lines of plants with such characteristics and improve their efficiency. The generation of genetic data related to phytoremediation needs to be managed through a database for long-term application such as storage, retrieval and modification when needed (Agarwal et al. 2020). Such database will help future researchers to understand and build on the technology for practical use. Research in phytoremediation has a corollary benefit in space biology. Several missions in Mars and Moon have excited scientists about the scope of colonizing other planets in space. This has resulted in identifying the factors that are needed for survival of humans in other planets. Among the many, one important factor is a continuous supply of food and to achieve this scientist have investigated on growing plants in space. Recent reports about the growth of Arabidopsis in moon soil and growing plants in the International Space Station have significantly encouraged the possibility of growing plants in space (Kiss et al. 2019; Khodadad et al. 2020; Paul et al. 2022). The soil composition in moon and mars are not as similar as found on earth and has elements in concentrations that are considered toxic in earth's atmosphere. Thus, the knowledge gained through the understanding of heavy metal tolerance in plants will essentially benefit space biology.

Besides plant biology, several disciplines of science such as chemistry, soil science, microbiology, biochemistry, genetics, ecology, economics, and chemical engineering can significantly contribute to the improvement of the phytoremediation technology. The "Omics" technologies such as genomics, metagenomics, transcriptomics, proteomics, metabolomics along with bioinformatics can be utilized for improvement of heavy metal phytoremediation through manipulation of membrane-bound transporters, translocation and compartmentation of ions, management of oxidative stress, and detoxification by caging of heavy metals to compounds and render them unavailable to the biological system. Though several plants capable of phytoremediation has been identified, however, the potential of phytoremediation in the large diversity of plant kingdom is largely unexplored. With the availability of sequencing technology and bioinformatics, such identification might be possible by investigating similarities in gene sequences and protein products. The technology of phytoremediation can be integrated with similar environmental and economic objectives such as carbon sequestration and biofuel production. The goal of heavy metal phytoremediation is their accumulation in plant parts for extraction, while the target of carbon sequestration is trapping the atmospheric carbon in plant tissues and biofuel projects are geared towards the accumulation of cellulosic organic mass in plants that can be utilized for biofuel production. Since all the above-mentioned three technologies have a common goal of environmental sustainability and to achieve that the improvement of plant biomass and understanding of element compartmentation is necessary. Identification of genetic traits for phytoremediation, carbon sequestration and biofuel production (De Deyn et al. 2008; Yadav et al. 2010; Jansson et al. 2010; Marmiroli et al. 2011; Wang et al. 2013) is the first step that can lead to metabolic engineering of plants with such properties.

Though the concept of phytoremediation is relatively recent and most of the work has been exploratory and superficial in nature besides being limited to laboratory and greenhouse. However, unlike the controlled environment in greenhouse, the field has several confounding factors such as soil nutrients, soil amendments, soil chemistry, soil particle size, water retention capacity, precipitation, temperature, wind, seasonal pathogens, and competition from other plants for limited resources. Such factors may interact, and it is important to estimate the effect of these factors on phytoremediation. Thus, more numerous and large-scale field trials of hyperaccumulators are important to estimate the confidence level before this technology can be patented and commercialized. Understanding the mechanism of heavy metal acquisition by plant roots is extremely important for phytoremediation research, however, above-ground plant tissues have been mostly investigated for the process. Innovation in technology (de Boon et al. 2022; Acharya 2022; Runck et al. 2022) can accelerate the investigation of the below-ground plant root tissues that has largely been ignored thus improving phytoremediation technology (Acharya 2023). The detailed understanding of the process may lead to their integration with agriculture that can lead to sustainable environment and better agriculture management practices (Acharya 2021; Mandal et al. 2022; Guo et al. 2023). Though this is an applied field of research, however, progress in this technology can also propel fundamental research on ion transport, compartmentation, and assimilation. The technology of heavy metal phytoremediation is at its early stage of development, however, holds immense potential for sustainable environment, agricultural practices, food security and research related to space biology.

## Acknowledgements

We thank the anonymous reviewers and editorial board for their support. This work was not supported by any funding such as external federal/state grants or internal university grants.

**Availability of data and material:** Not applicable.

**Competing interests:** The authors declare no competing interests.

**Funding:** No funding was obtained for this study.

**Authors' contribution:**

Aniruddha Acharya was involved in conceptualization, writing, and editing the manuscript,

Enrique Perez was involved in literature review, editing the manuscript, construction of tables and figures besides organizing the structure of the manuscript.

Miller Maddox-Mandolini was involved in editing the manuscript.

Hania De La Fuente was involved in literature review, editing the manuscript, construction of tables and figures besides organizing the structure of the manuscript.

| Elements | Plants | Concentrations of heavy metals | Mechanism | Type | References |
|---|---|---|---|---|---|
| As | *Pteris vittata* | 1639 mg Kg$^{-1}$ | Rhizofiltration | Hydroponic | Wu et al. 2009 |
| As | *Pteris vittata* | 1373mg Kg$^{-1}$ | Phytoextraction | Field | Wu et al. 2007 |
| As | *Azolla caroliniana* | 397 mg Kg$^{-1}$ | Rhizofiltration | Field | Favas et al. 2012 |
| Cd | *Phytolacca americana* | 637 mg Kg$^{-1}$ | Rhizofiltration | Hydroponic | Liu et al. 2010 |
| Cd | *Tagetes erecta* | 346 mg Kg$^{-1}$ | Phytoextraction | Pot | Madanan et al. 2021 |
| Cd | *Nicotiana tabacum* | 23 mg Kg$^{-1}$ | Phytoextraction | Field | Yang et al. 2019 |
| Co | *Nyssa sylvatica* | 438 mg Kg$^{-1}$ | Phytoextraction | Pot | McLeod & Ciravolo 2007 |
| Cr | *Leptochloa fusca* | 93 mg Kg$^{-1}$ | Phytostabilization | Pot | Ullah et al. 2021 |
| Cr | *Brachiaria mutica* | 18 mg Kg$^{-1}$ | Phytostabilization | Pot | Ullah et al. 2021 |
| Cu | *Khaya ivorensis* | 329 mg Kg$^{-1}$ | Phytoextraction | Pot | Covre et al. 2020 |
| Cu | *Commelina communis* | 1119 mg Kg$^{-1}$ | Rhizofiltration | Hydroponic | Wang et al. 2004 |
| Hg | *Helianthus tuberosus* | 1 mg Kg$^{-1}$ | Phytostabilization | Pot | Sas-Nowosielska et al. 2008 |
| Hg | *Pistia stratiotes* | 83 mg Kg$^{-1}$ | Rhizofiltration | Hydroponic | Skinner et al. 2007 |
| Mn | *Schima superba* | 62412 mg Kg$^{-1}$ | Phytoextraction | Pot | Yang et al. 2008 |
| Na | *Alternanthera philoxeroides* | 145000 mg Kg$^{-1}$ | Rhizofiltration | Hydroponic | Islam et al. 2019 |
| Ni | *Alyssum murale* | 443 mg Kg$^{-1}$ | Phytoextraction | Field | Tumi et al. 2012 |
| Ni | *Bidens pilosa* | 305 mg Kg$^{-1}$ | Phytostimulation | Pot | Liu et al. 2019 |

| Ni | *Leptoplax emarginata* | 7864 mg Kg$^{-1}$ | Phytoextraction | Field | Pardo et al. 2018 |
|---|---|---|---|---|---|
| Pb | *Sedum alfredii* | 915 mg Kg$^{-1}$ | Rhizofiltration | Hydroponic | He et al. 2003 |
| Pb | *Arundinaria argenteostriata* | 1117 mg Kg$^{-1}$ | Phytostimulation | Pot | Jiang et al. 2019 |
| Zn | *Thlaspi caerulescens* | 20000 mg Kg$^{-1}$ | Phytoextraction | Pot | Küpper et al. 1999 |
| Zn | *Coronopus didymus* | 1848 mg Kg$^{-1}$ | Phytoextraction | Pot | Sidhu et al. 2020 |

Table 1: Commonly used plants for investigation of heavy metal hyperaccumulation and/or phytoremediation capacity

| Different sources of Heavy Metals | Common Heavy Metal Candidates | Phytoremediation Categories | Factors influencing Phytoremediation |
|---|---|---|---|
| Mining of ores | As; Ag; Au Cr; Co; Cd; Cu; Fe; Hg; Mn; Ni; Pb; Se; Zn | Phytoaccumulation | Robust root system |
| Fossil fuel | | Phytominning | Bioavailability of elements |
| Volcanic activity | | Phytofiltration | Heavy metal tolerance capacity |
| Weathering of rocks | | Rhizofiltration | Growth rate of plant |
| Industrial waste | | Phytodesalination | Biomass accumulation |
| Fertilizers | | Phytostabilization | Ease of cultivation |
| Herbicides | | Phytovolatalization | Maintenance of plants |
| Pesticides | | Phytodegradation | Ease of Harvesting |
| Agriculture | | Phytotransformation | Ease of extraction of heavy metal from plant parts |
| Anthropogenic activity | | | Pathogens that can damage the plant |
| | | | Herbivores that can consume the plant parts |

Table 2: Bird's eye view of heavy metal phytoremediation.

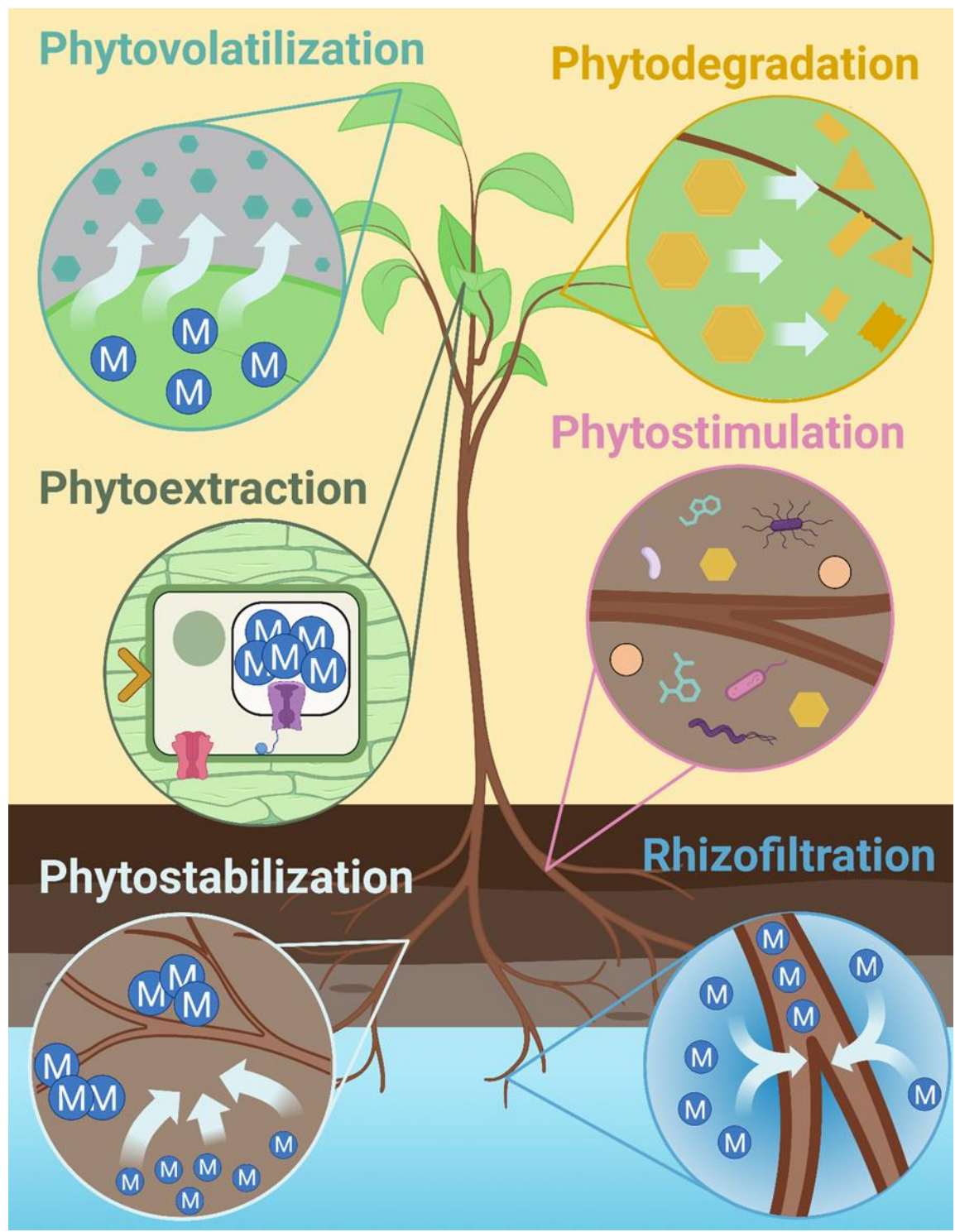

Image created by BioRender illustration software

Figure 1: The different mechanisms of phytoremediation of heavy metals that includes phytovolatilization, phytodegradation, phytoextraction, phytostimulation, phytostabilization and rhizofiltration. Though these processes are mechanistically different, however, they share a common target of recovering soil and water from pollutants.

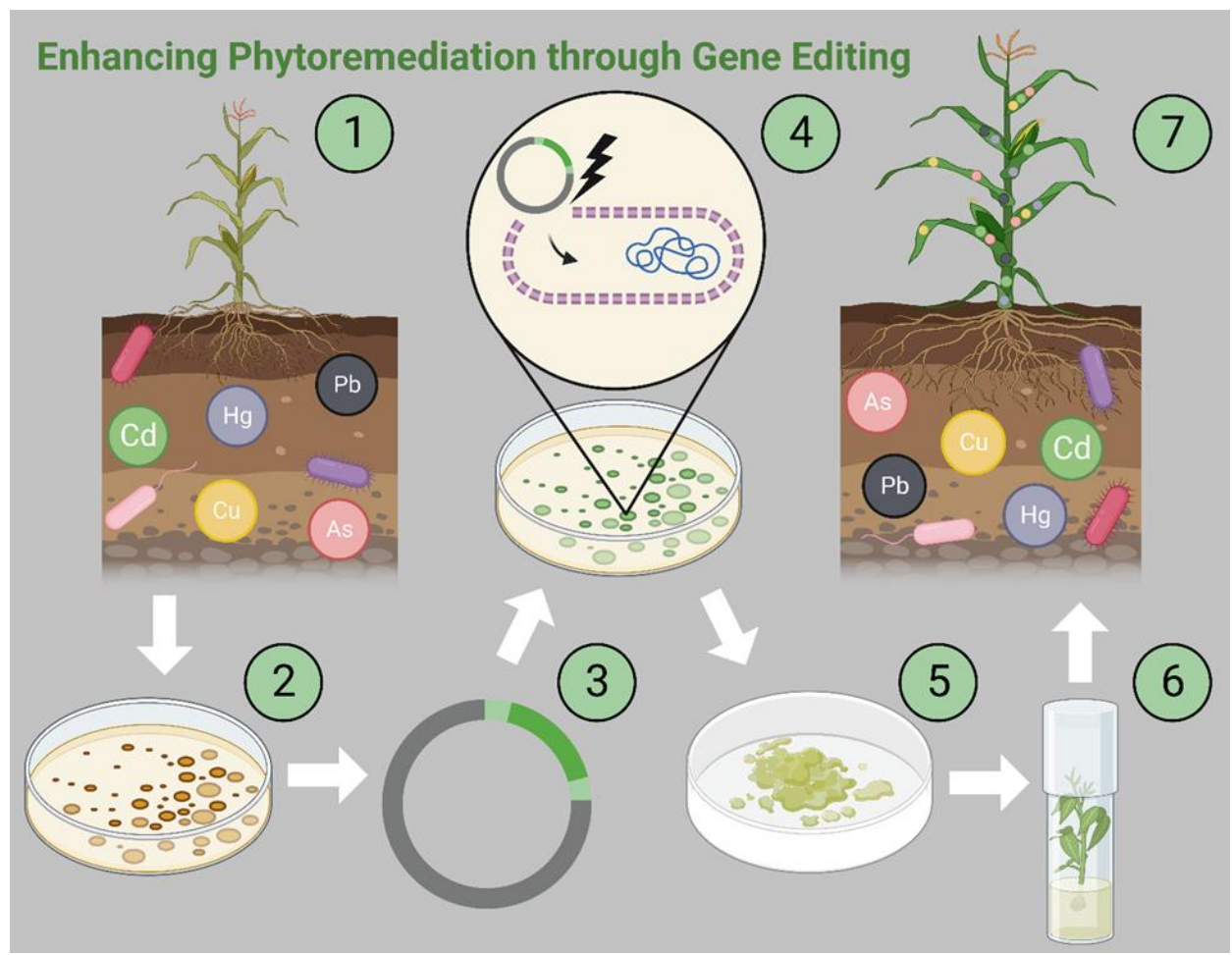

Image created by BioRender illustration software

Figure 2: Prospects of enhancing the phytoremediation technology through genetic engineering and genome editing where microbes and/or plant genes responsible for phytoremediation can be identified and cloned into a vector (1, 2&3) for genetic engineering. CRISPR gene editing technology can assist in precise gene manipulation to enhance the phytoremediation process. The gene of interest can be transferred to pluripotent callus cells through gene gun or *Agrobacterium* mediated transformation (4&5). Finally, the genetically modified plant can be regenerated in vitro and transferred to the field (6&7).

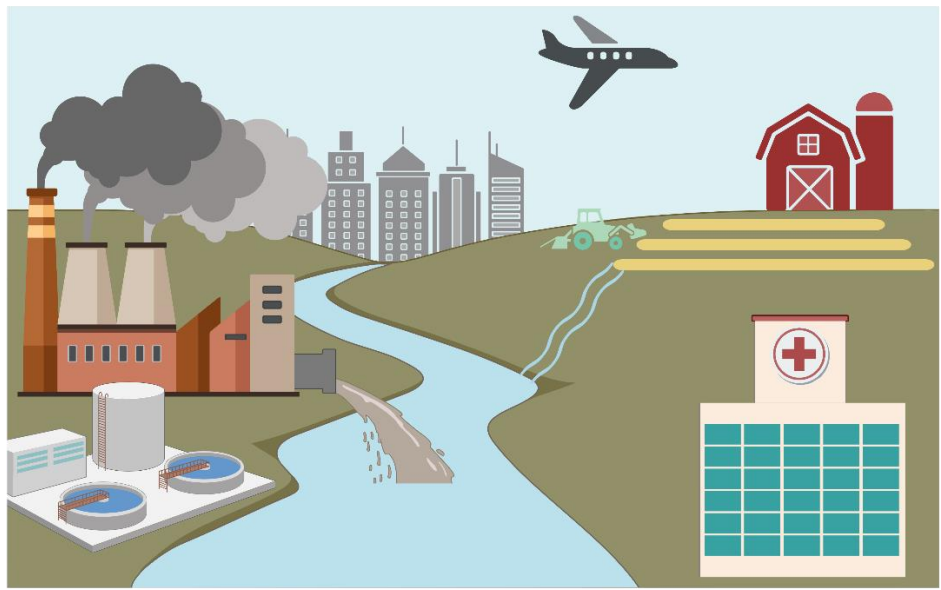

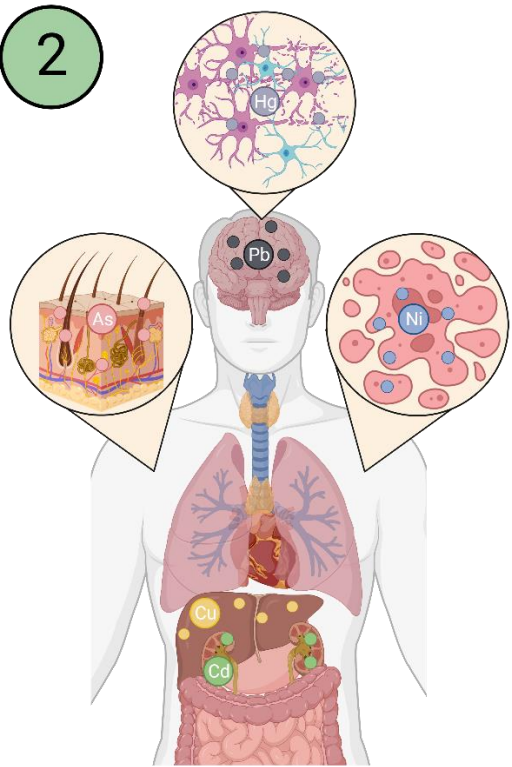

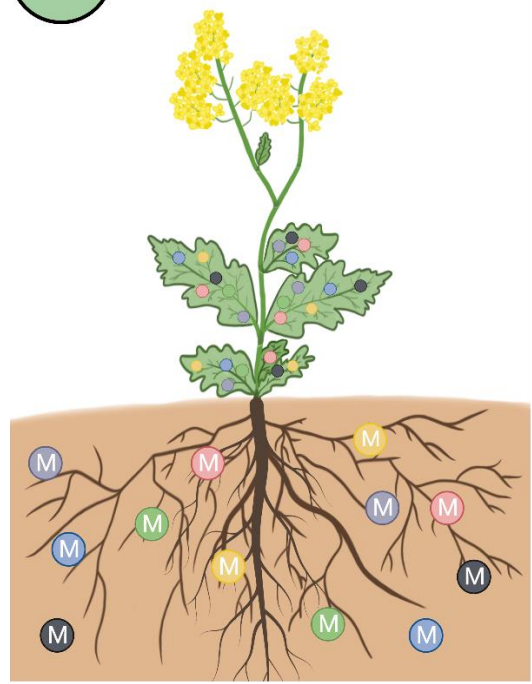

Image created by BioRender illustration software.

GRAPHICAL ABSTRACT